\documentclass[prd,twocolumn,showpacs]{revtex4}
\usepackage{hyperref,amssymb,amsmath,mathrsfs,bm,graphicx}
\usepackage{enumitem}
\usepackage{color}
\begin{document}
\title{Geodesics in the linearized multipole solution: Distinguishing black
holes from naked singularities} 
\author{J. L. Hernandez-Pastora}
\email{jlhp@usal.es}
\affiliation{Departamento de Matematica Aplicada and Instituto Universitario de
Fisica Fundamental y Matematicas, Universidad de Salamanca, Salamanca, Spain}
\author{L. Herrera}
\email{lherrera@usal.es}
\altaffiliation{Also at U.C.V., Caracas}
\affiliation{Departamento de F\'\i sica Te\' orica e Historia de la Ciencia,
Universidad del Pa\'\i s Vasco, Bilbao 48940, Spain}
\author{J. Ospino}
\email{j.ospino@usal.es}
\affiliation{Departamento de Matematica Aplicada, Universidad de Salamanca,
Salamanca, Spain}
\begin{abstract}
We analyze the behaviour of geodesic motion of test particles in the spacetime
of a specific class of axially symmetric static vacuum solutions to the Einstein
equations, hereafter referred to as  linearized multipole solution (LM). We
discuss about its suitability to describe a quasi-spherical spacetime. The
existence of an ISCO (innermost stable circular orbit)  very close to the
(singular) horizon  of the source, is established. The existence of such  stable
 orbit, inner than the one of the Schwarzschild metric, as well as the
appearance of a splitting in the admissible region of circular orbits, is shown
to be due  to  the multipole structure  of the solution, thereby providing
additional potential observational evidence for distinguishing Schwarzschild
black holes from naked singularities.
\end{abstract}
\date{\today}
\maketitle

\section{Introduction}

As it follows from the Israel theorem \cite{israel},  the only static and
asymptotically-flat vacuum space-time possessing a regular horizon is the
Schwarzschild solution. For all
the other Weyl exterior solutions \cite{weyl1, weyl2, weyl3, weyln, stephani},
the physical components of the Riemann tensor exhibit singularities at  the
infinite red shift surface. Even though we shall restrict ourselves to the
static case,  it is  worth noting  that a result similar to Israel theorem
exists for
stationary solutions with respect to the Kerr metric \cite{carter, hawk, wald}.

Now, sphericity is a common assumption in the description of compact objects,
where  deviations from spherical symmetry are likely to be incidental rather
than basic features of these systems.

Furthermore, if the field produced by a self--gravitating system is not
particularly intense (the boundary of the source is
much larger than the infinite redshift surface ) and fluctuations  off spherical symmetry are
slight, then there is no problem in representing the corresponding deviations
from spherical symmetry (both inside and outside the source) as a suitable
perturbation of the spherically symmetric exact solution \cite{Letelier}.

However, as the object becomes more
and more compact, such perturbative scheme will eventually fail close to the
source. Indeed, as is well known \cite{i2, i3, i1,i4, i5, in}, though usually
overlooked, as the boundary surface of the source approaches the infinite redshift surface, any finite
perturbation of the Schwarzschild spacetime, becomes fundamentally different
from the corresponding exact solution representing the quasi--spherical
spacetime,  even if the latter is characterized by parameters
whose values are arbitrarily close to those corresponding to Schwarzschild
metric. This in turn is just an expression of the Israel theorem.

In other words, for strong gravitational
fields, there exists a bifurcation between the perturbed Schwarzschild metric 
and all the
other Weyl metrics (in the case of gravitational perturbations), no matter how
small are the multipole moments (higher than monopole)
 of the source. Examples of such a  bifurcation have been brought out in the
study of the trajectories of test particles in the
$\gamma$ spacetime \cite{zipo, darmois, z, gau, coo, voo, espo, vir}, and  in
the M-Q spacetime \cite{mq1, mq2}, for orbits close to the infinite redshift surface 
\cite{HS},\cite{Herrera1}. 

Due to the bifurcation mentioned above, a fundamental question arises: How
should we describe the quasi--spherical space--time resulting from the
fluctuations off Schwarzschild?:
\begin{itemize}
\item (a) by means of a perturbed Schwarzschild metric producing a black hole

 or 

\item (b) by means of an exact solution to Einstein equations, whose
(radiatable) multipole moments are arbitrarily small, though non--vanishing, and
leading to a naked singularity ?
\end{itemize}

As we shall see here, the quandary above might be solved by comparing  the
behaviour of circular geodesics in either case.

Indeed, in spite of some results obtained in the study of the source of
quasi--spherical spacetimes
\cite{Herrera},
\cite{Herreramass}, which  favor the scenario (b), we are well aware of the fact
that, presently, most researchers, favor scenario (a). 
Nevertheless, the doubt remains, and the very different behaviour  of the system
implied by the bifurcation mentioned above, opens the way for proposing
observational scenarios allowing for distinguishing between black holes and
naked singularities. 
In fact this issue has attracted the attention of many researchers in recent
years (see \cite{dis1, dis2, dis3, dis4, dis5, dis6, dis7, dis8, dis9, dis10,
dis11, dis12, dis13, dis14} and references therein).

However  an important open question arises, related to the proposed approach,
namely: since there are as many
different (physically distinguishable) Weyl solutions as there are different
harmonic functions, which among Weyl solutions is the best
entitled to describe small deviations from spherical symmetry?

In the past  different authors have resorted to different metrics to describe
deviations from spherical symmetry; e.g.   the $\gamma$ metric or the M-Q
spacetime in \cite{HS, Herrera1, Herrera, Herreramass, dis13}, the Young-Coulter
solution \cite{young} in \cite{bini1},  the Quevedo-Mashhoon solution
\cite{mash} in \cite{bini2}, or the Manko-Novikov solution \cite{manko} in
\cite{dis14}. 

The rationale  behind the  choice of the $\gamma $-metric is based on  the fact
that it  corresponds to a solution of the
Laplace equation (in cylindrical coordinates) with a singularity
structure similar to that of the Schwarzschild solution (a line segment). In
this
sense the $\gamma $-metric appears as a ``natural'' generalization of
Schwarzschild space-time to the axisymmetric case. 

On the other hand, due to its relativistic multipole structure, the M--Q
solution (more exactly, a sub--class of this solution
M-Q$^{(1)}$
\cite{mq1}) may be interpreted as a quadrupole correction to the Schwarzschild
space--time, and therefore  represents a good candidate among known Weyl
solutions, to describe small
deviations from spherical symmetry.

However it should be obvious that the question above has not a unique answer
(there are an infinite number of ways of being non--spherical, so to speak) and
therefore in the study of
any specific problem, the choice of the corresponding Weyl spacetime has to be
reasoned.

In this work we intend to use yet another exact solution of the Weyl family, in
order to describe deviations from spherical symmetry. Such a solution is the so
called LM metric \cite{dumbel}, its properties and the reasons behind its choice
to describe a quasi--spherical  spacetime are presented  in the next section.
Next we shall calculate the circular geodesics in that spacetime an compare its
behaviour with the spherically symmetric case. The most relevant result
emerging from that analysis is the existence of stable 
innermost  circular orbits  very close to the
(singular) horizon  of the source, and  inner than the one of the Schwarzschild
metric.

\section{The LM spacetime}
As mentioned in the Introduction we shall carry out a study of circular
geodesics in the spacetime of the LM spacetime. Thus, we shall first very
briefly revise such a metric and provide arguments justifying its use to
describe quasi--spherical  (axially symmetric and static) spacetime (see
\cite{dumbel} for details). Finally we shall present the multipole structure of
the solution.
\subsection{The metric}

As is known, the line element of  a static and axisymmetric   vacuum space-time
is represented in Weyl form as follows
\begin{equation}
ds^2 = -e^{2\Psi} dt^2 +e^{-2\Psi}\left[
e^{2\gamma}\left(d\rho^2+dz^2\right)+\rho^2 d\varphi^2\right] \ ,
\label{weylds}
\end{equation}
where $\Psi$ and $\gamma$ are functions of the cylindrical coordinates $\rho$
and
$z$ alone. The metric function $\Psi$ is a solution of the Laplace's equation
($\triangle \Psi=0$), and the other metric function $\gamma$ satisfies a system
of differential equations whose integrability condition is just the equation for
the function $\Psi$. Thus, the Weyl family of solutions with  a good
asymptotical
behaviour is given in associated spherical Weyl coordinates $\{r, \theta\}$ as
\begin{equation}
  \Psi=\sum_{n=0}^\infty\frac{a_n}{r^{n+1}}P_n(\omega) \ ,
\label{psi}
\end{equation}
where  $r\equiv \sqrt{\rho^2+z^2}$, $\omega\equiv \cos \theta=z/r$ and $P_n$
denotes the Legendre polynomial.

Thus the line
element now reads
\begin{equation}
 ds^2= -e^{2\Psi} dt^2+e^{-2\Psi+2\gamma}
(dr^2+r^2d\theta^2)+e^{-2\Psi}r^2\sin^2\theta  d\varphi^2  .
\label{dsesferic}
\end{equation}

Also, as is well known, in spite of the form of (\ref{psi}), coefficients  $a_n$
are not the relativistic multipole moments (RMM) of the solution, as defined  
for static
and axisymmetric vacuum solutions by Geroch \cite{geroch1, geroch2} and Thorne
\cite{thorne}. However the ``Newtonian'' moments   $a_n$, which provide the so
called ``Newtonian image'' of the solution,  can be  expressed as functions of
the
RMM  \cite{fhp, sueco2, bakdal, sueco3}. Although the full  relations linking
both sets of coefficients
are extremely complicated, they can be used to obtain relatively simple formulas
for the coefficients $\{a_n\}$ in situations where the deviation of the
relativistic solution from spherical symmetry is small. This issue has been
discussed in some detail in \cite{tesis}, \cite{mq1}, \cite{sueco}.

A  solution of the Weyl family that
represents the exterior gravitational field of a mass distribution whose
multipole structure
 only possesses mass $M$ and Quadrupole moment $Q$ was found in  \cite{mq1}. 
This solution (M-Q)
has become a useful tool for describing small deviations from the spherically
symmetric
solution \cite{luisgyr, luisgeod, Herrera1, lhn2}. 

The basic idea underlying the obtention of   the
$M$-$Q$ solution is that $Q$ is small since we want to describe slight
deviations from  the
Schwarzschild solution, and all the RMM of higher order are
negligible. This assumption about the RMM higher than $Q$ is
supported by the
following argument: The Newtonian calculation
of the multipole moments of an ellipsoidal mass distribution
shows that as we move from lower to higher moments, their magnitudes decrease as
powers of the eccentricity of the ellipsoidal configuration (see \cite{tesis,
HBH} for details).
Then the M-Q solution is constructed as a
 sum of functions in a power series of the dimensionless quadrupole parameter
$q\equiv Q/M^3$ starting at the Schwarzschild solution as the first order, in
such a way that  the successive powers of $q$ control the desired corrections
to the spherical symmetry.

 The LM solution \cite{dumbel} was  constructed with the same purpose, namely:
to describe the gravitational field of a body slightly different from a sphere.
However the approach to find it, although similar,  is different from the one
used for the 	M-Q solution. In both cases, it should be clear that, describing 
non--spherical spacetimes, their physical components of the Riemann tensor 
exhibit singularities at the infinite redshift surface.

The rationale behind the LM solution is the following:  When attempting  to
describe an isolated compact body which is not spherically symmetric,
 all the RMM appear no matter how small is the deviation from spherical
symmetry.
Therefore, let us consider that all RMM appear in the solution that we want to
construct but let us restrict their magnitudes to be very small, so
that  we can neglect all terms in the Weyl coefficients whenever a cross product
of RMM's is
involved. This is the origin of  the name of the family of solutions: {\it
Linearized Multipole} (LM) solution. It should be observed that due to the
linearity of the Laplace equation, the so obtained solution is an exact solution
to Einstein equations.

Now, the expression for the metric function of the Weyl family solution
endowed with $g+1$ independent RMM, (the $LM$-solution),
 can be written, in prolate spheroidal coordinates \cite{erq}, as
follows
\begin{equation}
 \Psi=-\frac{H x}{x^2-y^2}-\sum_{n=0}^gQ_{2n}(x)P_{2n}(y) \left[ \sum_{j=n}^gH_j
C_{2j,2n} \right] ,
\label{psiLM}
\end{equation}
where $C_{n,k}$ are the coefficients appearing at the series expansion   of any
variable as a linear combination
of Legendre polynomials in that variable i.e., ${\displaystyle
\xi^n=\sum_{k=0}^{\infty} C_{n,k}P_k(\xi)}$), and 
\begin{equation}
 H\equiv \sum_{k=0}^gm_{2k}h(k) \qquad , \qquad H_j\equiv
\sum_{k=j}^gm_{2k}h_j(k) \ ,
\label{haches}
\end{equation}
where the parameter $m_{2k}\equiv\frac{M_{2k}}{M^{2k+1}}$ denotes the
dimensionless relativistic multipole moment of order $2^{2k}$-pole ($M_{2k}$), 
whereas  the explicit expressions for the coefficients $h_j(k)$ ($\forall
\ k
\geq j$, since $h_j(k)=0$ for $k<j$) and $h(k)$, are:
\begin{widetext}
\begin{equation}
h_j(k)=\frac{1}{2^{4k-1}}(-1)^{k-j-2}\displaystyle{{{4k+1} \choose
{2k}}}\frac{k^2+k/2-j}{(k+1)}\frac{(2k+2j)!}{(2j)!(k+j)!(k-j)!} \ , \quad  
h(k)= \frac{1}{2^{2k} (k+1)} {{4k+1}
\choose {2k}}  \ , \forall \ k>0 \ .
\label{hachek}
\end{equation}
\end{widetext}
The parameter $H$ can be calculated in terms of the coefficients $H_j$ as
follows:
\begin{equation}
 {\displaystyle H= \sum_{k=0}^g \frac{m_{2k}}{2^{2k} (k+1)} {{4k+1}
\choose {2k}} }=\left[1-\sum_{j=0}^g\frac{H_j}{2j+1}\right] \ .
\label{masabolanu}
\end{equation}

\vspace{4mm}

In terms of  its ``Newtonian image", the LM solution can be described by means
of an ``object-image'' whose
Newtonian gravitational potential (the gravitational potential corresponding to
the Newtonian image, not to confound with the weak field limit of the solution),
as well as 
its Newtonian multipole moments equal the metric function of the solution and
the Weyl coefficients respectively. That ``object-image'' is  represented by  a
kind of 
``dumbbell'' consisting of  a bar of length $2M$ with linear density $\mu$
given by an even  polynomial of degree $2g$ and two balls at each end of the bar
with mass $\nu$ (see \cite{dumbel} for details).

\vspace{4mm}

The other metric function $\gamma$ of the line element (\ref{weylds}) can be
obtained from the metric function $\Psi$ by solving
 the corresponding field equations
\begin{equation}
 \gamma_{\rho}=\rho(\Psi^2_{\rho}-\Psi^2_z) \quad \ , \quad \ \gamma_{z}=2 \rho
\Psi_{\rho} \Psi_z \ .
\end{equation}
There already exists an expression for this metric function \cite{erq} in terms
of the Weyl
coefficients of the series $\Psi$ (\ref{psi}), but it is
highly complicated to handle and a summation of a series is required to obtain
the analytic expression of the metric function.
One advantage inherent to  the dumbbell description of the solution consists of
an integral
expression (\cite{teixeira}) for the metric function $\gamma$ in terms of the
density
 of the dumbbell :
\begin{widetext}
\begin{equation}
 \gamma=\int_{-1}^1dX\int_{-1}^1dY \frac{\mu^d(X)\mu^d(Y)}{(Y-X)^2}
\left[\frac{r^2(1-\omega^2)+(r\omega-XM)(r\omega-YM)}{R(X)R(Y)}-1\right]  \ ,
\end{equation}
\end{widetext}
or equivalently,
\begin{equation}
\gamma=-M^2 r^2(1-\omega^2) \left[ \nu^2 A(r,\omega) +2 \nu
(I_-+I_+)+II\right] \ , 
\end{equation}
where  the following notation is used $R(X)\equiv \sqrt{r^2+X^2M^2-2r\omega
XM}$, $\mu^d(X)$ represents the density of the dumbbell and
\begin{widetext}
\begin{eqnarray}
&II \equiv {\displaystyle \int\int_{-1}^1 \frac{\mu(X)\mu(Y) \ dX
dY}{R(X)R(Y)\left[r^2(1-\omega^2)+(r\omega-XM)(r\omega-YM)+R(X)R(Y)\right] }
}\nonumber \\                                                                               
&I_{\pm} \equiv {\displaystyle \int_{-1}^1dX
\frac{\mu(X)}{r_{\pm}R(X)\left[r^2(1-\omega^2)+(r\omega-XM)(r\omega\pm
M)+r_{\pm}R(X)\right]}} \nonumber \\
&A(r,\omega)\equiv {\displaystyle
\frac{1}{4r_-^4}+\frac{1}{4r_+^4}+\frac{1}{r_-r_+\left[r^2-M^2+r_+r_-\right]}} \
.
\end{eqnarray}
\end{widetext}

The fact that the Newtonian image of the LM solution is a dumbbell is quite
convenient, since many properties of the solution can be described in  terms of
the density of the bar. Thus for example, it can be shown that the source of the
solution will be prolate (oblate) if $\mu$ is smaller (greater) than $1/2$. 
Also, as we shall see  below, the possible existence of an ISCO, inner to the
one correspondig to the spherically symmetric case is related to a condition on
$\mu$ at the origin  (\ref{limite}). 

\subsection{Mutipole structure of the solution}

As already mentioned, the RMM of a Weyl solution can be calculated  in terms of
the coefficients $a_n$. This relation can be inverted to
obtain the Newtonian moments ($a_n$) in terms of the RMM.
 The assumption used to construct the LM solution is that every RMM is small,
implying that we neglect all the terms with coupling interaction between RMM
appearing in the Weyl coefficients $a_n$ (once again it should be emphasized
that, due to the linearity of the Laplace equation, the so obtained metric is an
exact solution to Einstein equations). With this selection of the
coefficients we can consider that the solution possesses a finite number of
parameters ($q\equiv m_2$, $m_{2i}$, with $1< i \leq g$) that  represent each
RMM of the solution. 

Thus, the first RMM of the solution are the following (odd moments are null
because of the equatorial symmetry):
\begin{eqnarray}
M_0&=&M \nonumber\\
M_2&=&M^3 q \nonumber\\
M_4&=&M^5 m_4 \nonumber\\
M_6&=&M^7 \left(m_6-\frac{60}{77}q^2\right) \nonumber\\
M_8&=&M^9 \left(m_8-\frac{226}{143}q
m_4-\frac{1060}{3003}q^2-\frac{40}{143}q^3\right) \nonumber\\
M_{10}&=&M^{11} \left(m_{10}-\frac{28616}{46189}q m_4-\frac{566}{323}q
m_6-\frac{30870}{46189}m_4^2\right.\nonumber\\
&-& \left.\frac{19880}{138567}q^2-\frac{39150}{46189}q^2
m_4+\frac{146500}{323323}q^3\right) \ .
\end{eqnarray}

\section{Geodesics}

We shall now study the geodesic motion of test particles in the spacetime of the
LM solution. We shall restric ourselves to the case of geodesics with constant
$\theta$ and
$\displaystyle{\frac{d\varphi}{d\sigma}\neq 0}$, i.e.,  those constrained to a
constant hypersurface ($\theta=\theta_0$)
with coordinates $\{t,r,\varphi\}$.

Therefore, we obtain on the equatorial
plane the following expression (see \cite{luisgeod}):
\begin{equation}
 \left(\frac{dr}{d\sigma}\right)^2+V_{eff}=C \label{ocho} \ ,
\end{equation}
where $\sigma$ denotes the affine parameter along the geodesic, $C$ is a
constant and
$V_{eff}$ is an effective potential which can be obtained  by
integration as follows
\begin{equation}
 V_{eff}=\int \frac{k}{g_{11}} \partial_r \ln\left(\frac{g_{11}}{k}\right) dr =
-\frac{k}{g_{11}} \
.\label{eff}
\end{equation}
with $k\equiv\displaystyle{\epsilon-\frac{h^2}{g_{00}}-\frac{l^2}{g_{33}}}$,
where $h$ and $l$ represent the energy and angular momentum per unit mass
respectively, and $\epsilon$ denotes the norm of the tangent vector to the
geodesic $z^\alpha$.

From Eqs. (\ref{ocho}) and (\ref{eff}) we have that
\begin{equation}
 \left(\frac{du}{d\varphi}\right)^2=\frac{k}{g_{11}}\frac{g_{33}^2}{l^2} u^4 \ ,
\label{dudr}
\end{equation}
where  $u\equiv 1/r$.

The above equations, lead, for the line element  (\ref{weylds}), to 
\begin{equation}
\left(\frac{du}{d\varphi}\right)^2=\frac{F(r)}{l^2 e^{2\gamma+4\Psi}},
\label{geod}
\end{equation}
\begin{equation} V_{eff}=-\frac{F(r)}{e^{2\gamma}}, \label{poten}
\end{equation}
where the function $F(r)\equiv k e^{2\Psi}$ for timelike geodesics on the
equatorial plane is
\begin{equation}
 F(r)=-e^{2\Psi}+h^2-\frac{l^2}{r^2}e^{4\Psi}.
\label{efe}
\end{equation}

When looking for circular orbits we search for the stationary solutions of the
autonomous partial differential equation (\ref{geod}), i.e., 
$\displaystyle{ \frac{du}{dr}=0 \Longleftrightarrow u=cte}$. Hence we can say
that the circular orbits are defined by radial values $r=R_i$ where the
following condition is satisfied:
\begin{equation}
 F(R_i)=\frac{dF}{dr}(R_i)=0 \ ,
\label{ffpri}
\end{equation}
since the extremals of the effective potential satisfy (prime denotes derivative
with respect to $r$)
\begin{eqnarray}
 \frac{dV_{eff}}{dr}(R_i)=0=&\left(-F^{\prime}(R_i)+F(R_i) 2
\gamma^{\prime}(R_i)\right)e^{-2\gamma(R_i)} \nonumber\\
& \Rightarrow  F^{\prime}\equiv\frac{dF}{dr}(R_i)=0 \ .
\label{ffpri2}
\end{eqnarray}

Therefore, the circular orbits can be calculated by means of the function $F(r)$
without using the second metric function $\gamma$ since the complete
effective potential is not needed. 

The timelike geodesics described by a
pointlike particle around a circular orbit is defined by the zeros of 
both the function $F(r)$ and its derivative. The orbit $r=R_i$ is stable
($\displaystyle{\frac{d^2V_{eff}}{dr^2}>0}$) if $-F^{\prime \prime}(R_i)>0$ (the
minimum) and it is unstable ($\displaystyle{\frac{d^2V_{eff}}{dr^2}<0}$) if
$-F^{\prime \prime}(R_i)<0$ (the maximum). In the above it has been used that 
\begin{equation}
\frac{d^2V_{eff}}{dr^2}=e^{-2\gamma}\left(-F^{\prime\prime}+F^{
\prime}4\gamma^{\prime}+F
2\gamma^{\prime\prime}-(2\gamma^{\prime})^2F\right)\label{f1}
\end{equation}
and hence 
\begin{equation}
\frac{d^2V_{eff}}{dr^2}(R_i)=-e^{-2\gamma(R_i)}
F^{\prime\prime}(R_i) \ .
 \label{f2}
\end{equation}
Observe that the  specific energy of the geodesic orbit  is $E= -z_0$,
therefore, since $z_0=h=\frac{dt}{d \sigma} g_{00} <0$,  the parameter $h$ 
(with $V_{eff}=0$  or equivalently $F(R_i)=0$)  denotes,  up to a sign,  the
energy per unit of mass of the test particle  and it is fixed once the
extremals ($R_i$) of $F(r)$  are determined:
\begin{equation}
 h^2=e^{2\Psi(R_i)}\left[1+\frac{l^2}{R^2_i}e^{2\Psi(R_i)}\right].
\end{equation}
Also, observe that the conditions for circular orbits (\ref{ffpri}),
(\ref{ffpri2})
determine the values of $h$ and $l$ as follows:
\begin{equation}
 l^2_i=\left. \frac{r^3 \Psi^{\prime}}{e^{2\Psi} (1-2 r
\Psi^{\prime})}\right\rvert_{r=R_i} \ , \
h^2_i=\left. e^{2\Psi} \frac{1- r \Psi^{\prime}}{1-2 r
\Psi^{\prime}}\right\rvert_{r=R_i}
\end{equation}
Then, these parameters are constants of motion for each value of the radial coordinate $r=R_i$. In what follows we shall consider
 the angular parameter as a function of the radial coordinate $r$ for different circular orbits and hence we introduce 
 the notation 
\begin{equation}
L=L(r)\equiv {\displaystyle \frac{l^2}{4M^2}=\frac{r^3 \Psi^{\prime}}{4 M^2 e^{2\Psi} (1-2 r \Psi^{\prime})}} \ .
\label{lyh}
\end{equation}

\subsection{The spherically symmetric solution}

For the forthcoming discussion, it would be convenient to recover the
Schwarzschild  case, which is well  known. 

Depending on the value of  $L$ the function $-F(r)$ 
acquires a maximum and a minimum starting  from the particular value $L=3$
for
which both extremals coincide at $r_s=6M$. For large values of $L$ the
minimum
goes away asymptotically along $r_s/M=3$. In what follows the notation
$\lambda\equiv r_s/M$ shall be used, where $r_s$ denotes the radial
Schwarzschild
coordinate and it is related to the radial Weyl coordinate $r$ (on the
equatorial
plane) 
as follows:

\begin{equation}
 s\equiv \frac rM=\frac{r_s}{M}\sqrt{1-2 \frac{M}{r_s}} \ .
\label{rela}
\end{equation}
In figure 1a we plot the parameter $L$  as function  of the dimensionless
Schwarzschild radial  coordinate $r_s/M$, where the fact that   for the circular
orbits of  the Schwarzschild space-time: $r_s/M=2 L \left(1\pm \sqrt{1-3/
L}\right)$, has been used.   The value of the
parameter $h$ is taken to be zero, since it only generates a displacement of the
graphic along the vertical axis.  There exist 
certain value $h$ for each  extremal $R_i$, where $F(R_i)=0$.  In figure 1b, $-F(r)$  with its  extremals are shown for different
values of $L$.

\begin{figure}[h]
$$
\begin{array}{cc}
 \includegraphics[scale=0.21]{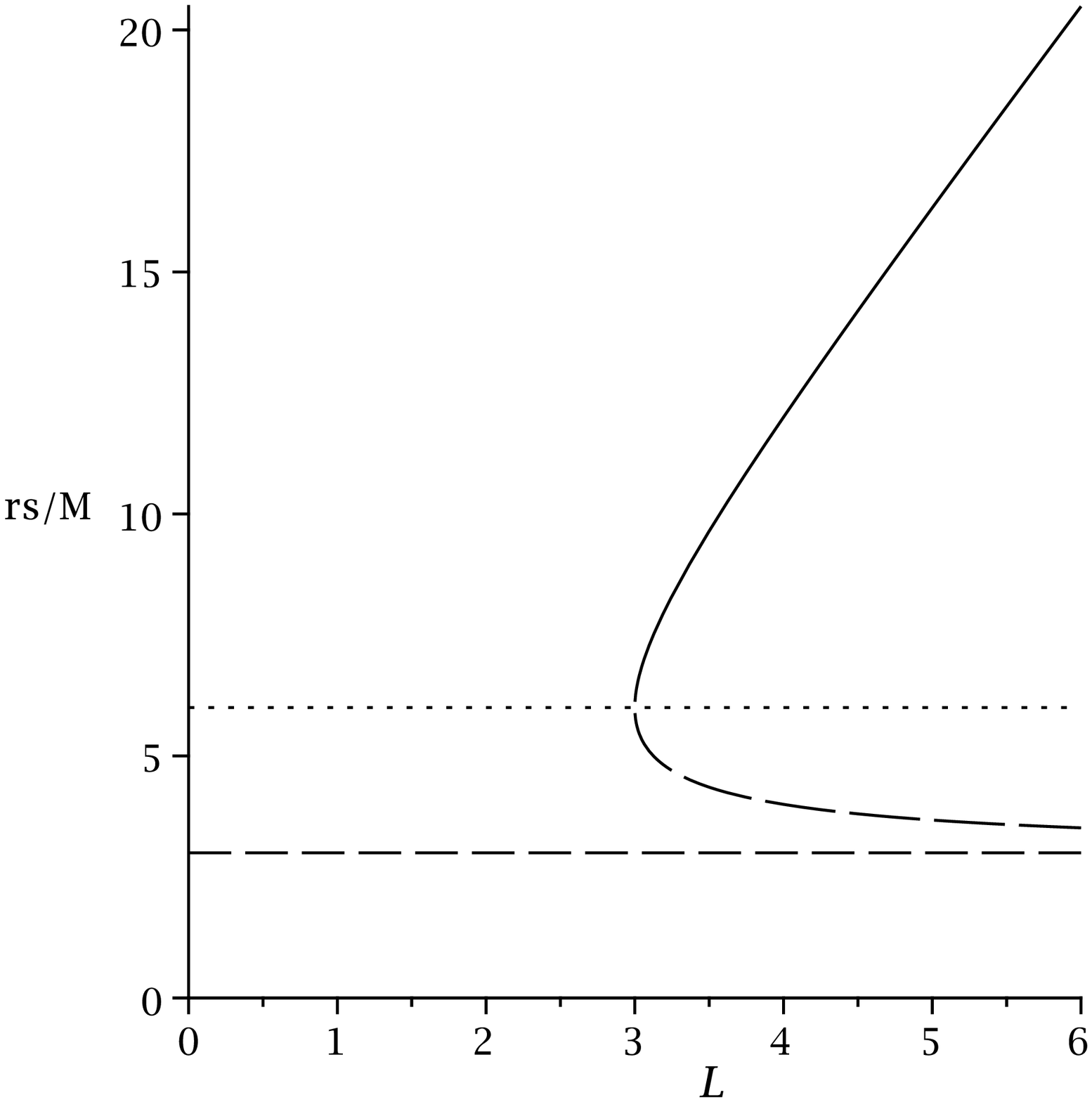}&
\includegraphics[scale=0.21]{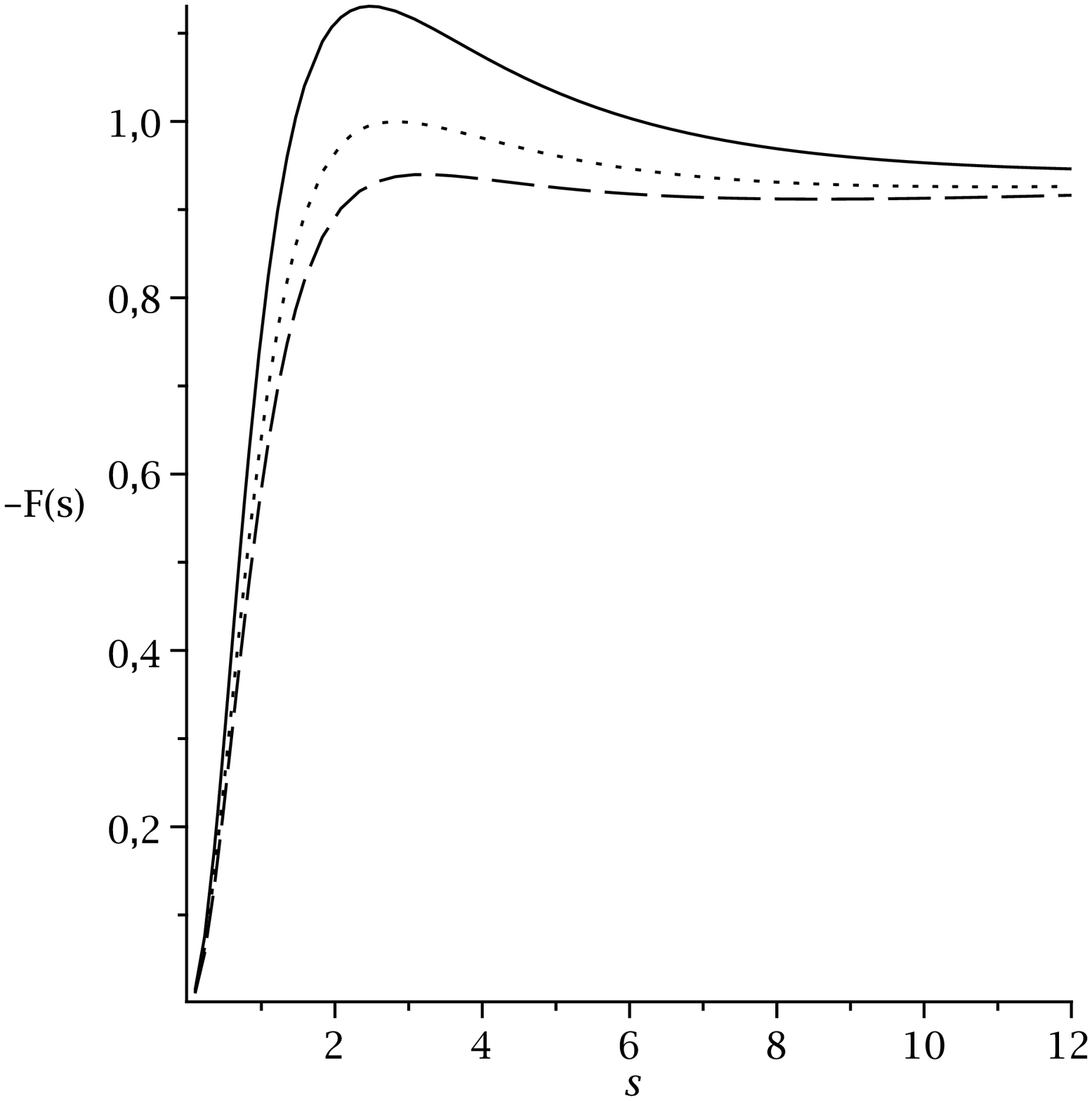}  \nonumber\\
(a) & (b) \nonumber
\end{array}
$$
\caption{\label{schwar} \it (a) The plot of  the parameter $L$   in terms of
the dimensionless Schwarzschild radial  coordinate $r_s/M$. For each value of the parameter $L$ (horizontal axis) the solid  line  and
the   long-dashed  line  provide the points where  the function $-F(s)$ acquires
 the  minimum or the maximum respectively. These values define the corresponding
radii of  the circular stable  or unstable orbits respectively. As is known, no
matter how large the  parameter $L$ would  be, the inner unstable orbit is
located at $r_s=3M$ (dashed horizontal line), and the bifurcation point between 
stable and unstable orbits (the marginally stable orbit) is located at $r_s=6M$
where the inner stable orbit is reached (shown with a  dot  line in the
graphic).  (b) In this
graphic the function $-F(s)$ is
represented for different values of $L$.  Starting from the solid line and
downward, the values  of the parameter $L$ are: $L= 5, 4, 3.5$. Let us
note that the value  $L=3$ corresponds to the marginally stable orbit, where
$F(s)$ has no extremals points but an inflection point.}
\end{figure}

\subsection{The LM solution}
Let us now analyze the situation in the LM solution. To derive the consequences
implied by the extremal condition $F{^\prime}(r)=0$, we need to solve 
numerically the
following transcendent equation
\begin{equation}
 e^{2\Psi}=\frac{r^3}{l^2}\frac{\Psi^{\prime}}{1-2r\Psi^{\prime}}.
\end{equation}
Nevertheless, a relevant information can be extracted from the analytical study
of the function $-F(r)$ (\ref{efe}). The calculation of that 
function for the LM solution yields 
\begin{equation}
 -F(s)= G^{C(s)}e^{A(s)}+\frac{4 L}{(\sqrt{s^2+1}+1)^2}
G^{2C(s)-1}e^{2A(s)}-h^2
\end{equation}
with the notation (for the case $g=2$)
\begin{eqnarray}
 {\displaystyle \ A(s) \equiv
-\sqrt{s^2+1}
B(s)-\frac{2H}{\sqrt{s^2+1}}}\\ 
{\displaystyle G\equiv\frac{\sqrt{s^2+1}-1}{\sqrt{s^2+1}+1}\ , \ 
C(s)\equiv
H_0-\frac 12 H_1 s^2+\frac 38 H_2 s^4}\\ 
{\displaystyle B(s)\equiv H_1+H_2\left(\frac 12-\frac 34 s^2\right)} \ .
\end{eqnarray}
These expressions are easily obtained from the metric function $\Psi$
(\ref{psiLM}) by considering it on the equatorial plane ($y=0$) and taking into
account that 
$x=\lambda -1=\sqrt{1+s^2}$ (the explicit expression of $\Psi$ in Weyl
coordinates can be seen in \cite{dumbel}). Let us note that $e^{2 \Psi}=G^{C(s)}
e^{A(s)}$ and $G^{2C(s)}={\displaystyle \frac{G^{2C(s)-1}
s^2}{(\sqrt{s^2+1}-1)^2}}$.

As can be seen in figure 2, the behaviour of the function $-F(r)$ is different
from that corresponding to the spherically symmetric case. 

Indeed, as   shown in that figure,  for certain values of the multipole
parameters $q\equiv m_{2}$ and $m_{4}$, the curve clearly shows a minimum close
to  the origin. This implies the existence of  an ISCO, inner to the one
corresponding to the spherically symmetric case, and therefore related to the
presence of the multipole moments ($m_2$ and $m_4$). Let us remember
 that $r=0$ corresponds to 
the infinite redshift surface  (\ref{rela}).

\begin{figure}[h]
\caption{\label{LMF}\it  The function $-F(s)$ is represented for the LM solution
possessing Monopole, quadrupole and $2^4$-pole moment ($h=0$ is considered). }
\includegraphics[scale=0.30]{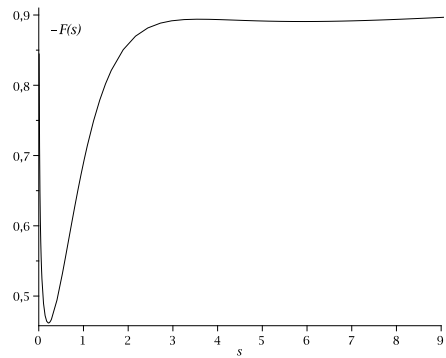}
\end{figure}

If we calculate the behaviour of the function $-F(s)$ when approaching the
horizon, we see that this 
function goes to infinity for certain values of the multipole parameters,
thereby exhibiting  the existence of a minimum which is
absent in the Schwarzschild solution:
\begin{equation}
\lim_{s\rightarrow 0} (-F(s))=
\left\{
\begin{array}{cc}
0 \quad , & \quad 2H_0-1>0 \\
\infty \quad , & \quad 2H_0-1<0 
\end{array}
\right. \ ,
\label{limite}
\end{equation}
since
$
\lim_{s\rightarrow 0} e^{2\Psi}=\lim_{s\rightarrow 0} G^{C(s)} e^A =0
$ because  $C(0)=H_0\equiv 2\mu^{LM}(0) >0$ (the density of the dumbbell bar  is
positive definite).

For the case of M-Q$^{(1)}$ solution, i.e. the LM solution with monopole and 
quadrupole moments alone, the existence of the ISCO is determined by
the following range of values of  the quadrupole parameter: $q\in
\left[\frac{4}{15}, \frac{8}{15}\right]$.

For the case of LM solution with monopole, quadrupole and $2^4$-pole
moments, the existence of the ISCO is determined by
the following range of values of  the quadrupole parameter: 

\begin{equation}
q \in
\left\{\nonumber
\begin{array}{cc}
\left[-\frac{32}{255}, -\frac{16}{255}\right], & \,    m_4=q\nonumber\\
 & \nonumber \\
\left[\frac{16}{375}, \frac{32}{375}\right], & \,  m_4=-q \nonumber
\end{array}
\right. \ ,
\label{iscom4q}
\end{equation}
 where we have assumed that the absolute value of both multipole moments are
identical (see (\cite{dumbel}) for details). The determination  of these ranges
of values is obtained from imposing two conditions: the positive definite
density condition and $2H_0-1<0$, that leads to $0<H_0<1/2$  (see the
figure 3 for details and a graphical characterization of these ranges).

\begin{figure}[h]
\caption{\label{qm4ranges}\it The domain of the existence of ISCO in the LM
solution is shown. Dotted  lines draw the condition assumed on the parameters $q$
and $m_4$  which are supposed to be  of equal magnitude (in absolute value). The
continuous line represents the limit (\ref{limite})
$1-\frac{15}{4}q+\frac{315}{16}m_4<0$ for the existence of ISCO, hence the
values  of $q$ must be situated on top of this line. The intersections of these
lines determine the upper value of $q$ if  it is negative or the lower bound if
it is positive, whereas the other extremes of the ranges are determined by the
definite-positive condition of the density (horizontal  dashed line and
dot-dashed line).}
\includegraphics[scale=0.20]{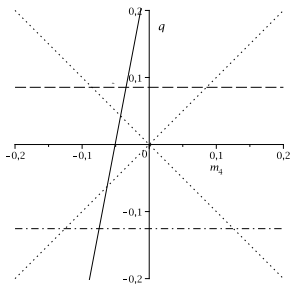}
\end{figure}

In addition, a  more relevant feature of these solutions is  obtained from the
study
of the marginally stable orbit (mso). Indeed, as is known,  for a circular (mso)
orbit, 
the angular parameter (as well as the energy) have extremal values. This
condition is just equal to $F^{\prime \prime}=0$ as 
can be seen by taking the derivative of equation (\ref{lyh})
\begin{equation}
 \frac{d L}{d r} = 0 \Longleftrightarrow 0= r \Psi^{\prime \prime}+3
\Psi^{\prime}+4 r^2
(\Psi^{\prime})^3-6r (\Psi^{\prime})^2 \ .
\end{equation}
Therefore, the circular equatorial motion is known to be stable when
$L^{\prime}>0$ and unstable for $L^{\prime}<0$. 
Let us notice that the epicyclic 
frequency is proportional to $L^{\prime}$, and hence the mso
($L^{\prime}=0$) determines the orbit with non horizontal oscillations.
The existence of an ISCO, as we have previously shown, can be confirmed when
studying the behaviour of $L$ in terms of  the orbital radius.

\begin{figure}[h]
\caption{\label{msotrio}\it  Localization of circular orbital radii in terms of
the parameter $L$ (in the vertical axis) for different values of the
quadrupolar parameter for the M-Q$^{(1)}$ solution. (a) For $q \in
\left(q_c,8/15\right]$. Starting from the solid line and downwards the values of
the corresponding quadrupolar parameter are $q=0.4, 0.42, 0.46$.  (b) For $q \in
\left(0,4/15\right]$. This curve corresponds to $q=0.2$ where the asymptota is
located at $\lambda=2.8125$ (a  smaller  value  than the corresponding for the 
Schwarzshild case $\lambda=3$). (c)  For $q \in \left(4/15 ,q_c\right]$.  This
piece wise curve corresponds to $q=0.3$ where the asymptotic lines bounding the
forbidden region are located at $\lambda_m=2.0473$ and $\lambda_M=2.6646$.}
$$
\begin{array}{cc}
(a) & \includegraphics[scale=0.20]{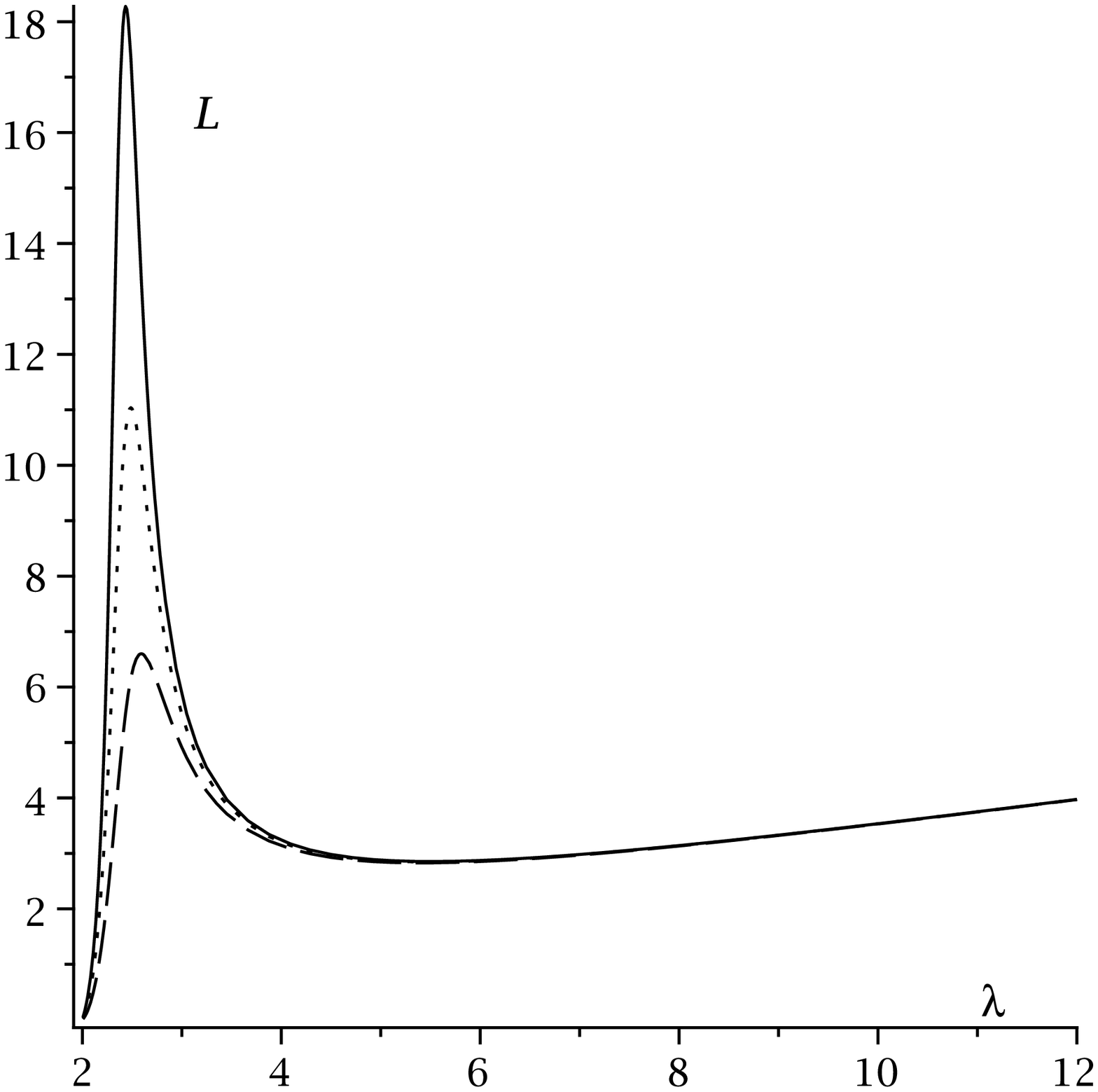}\\ \nonumber
(b) & \includegraphics[scale=0.20]{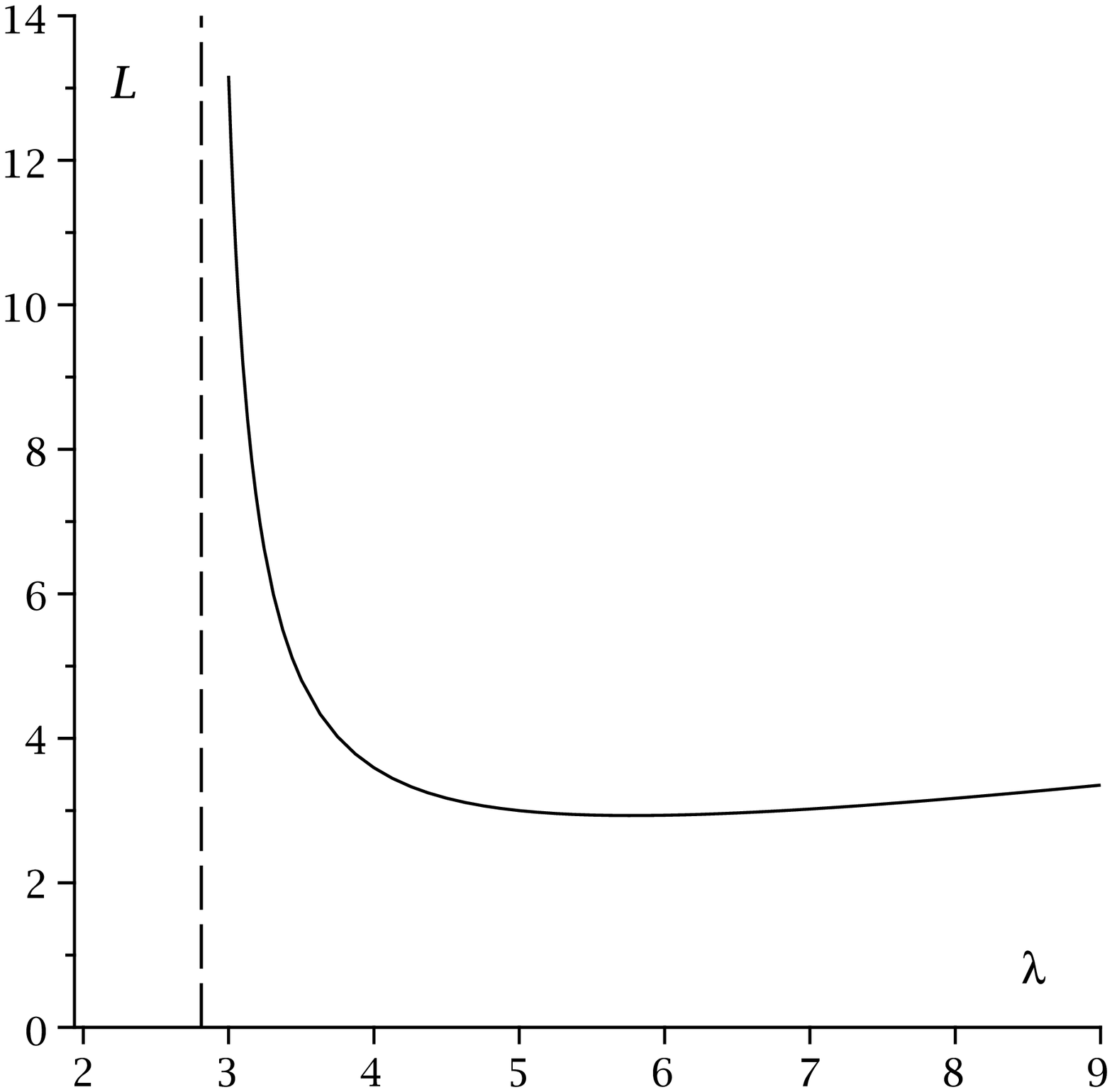}\\ \nonumber
(c) & \includegraphics[scale=0.20]{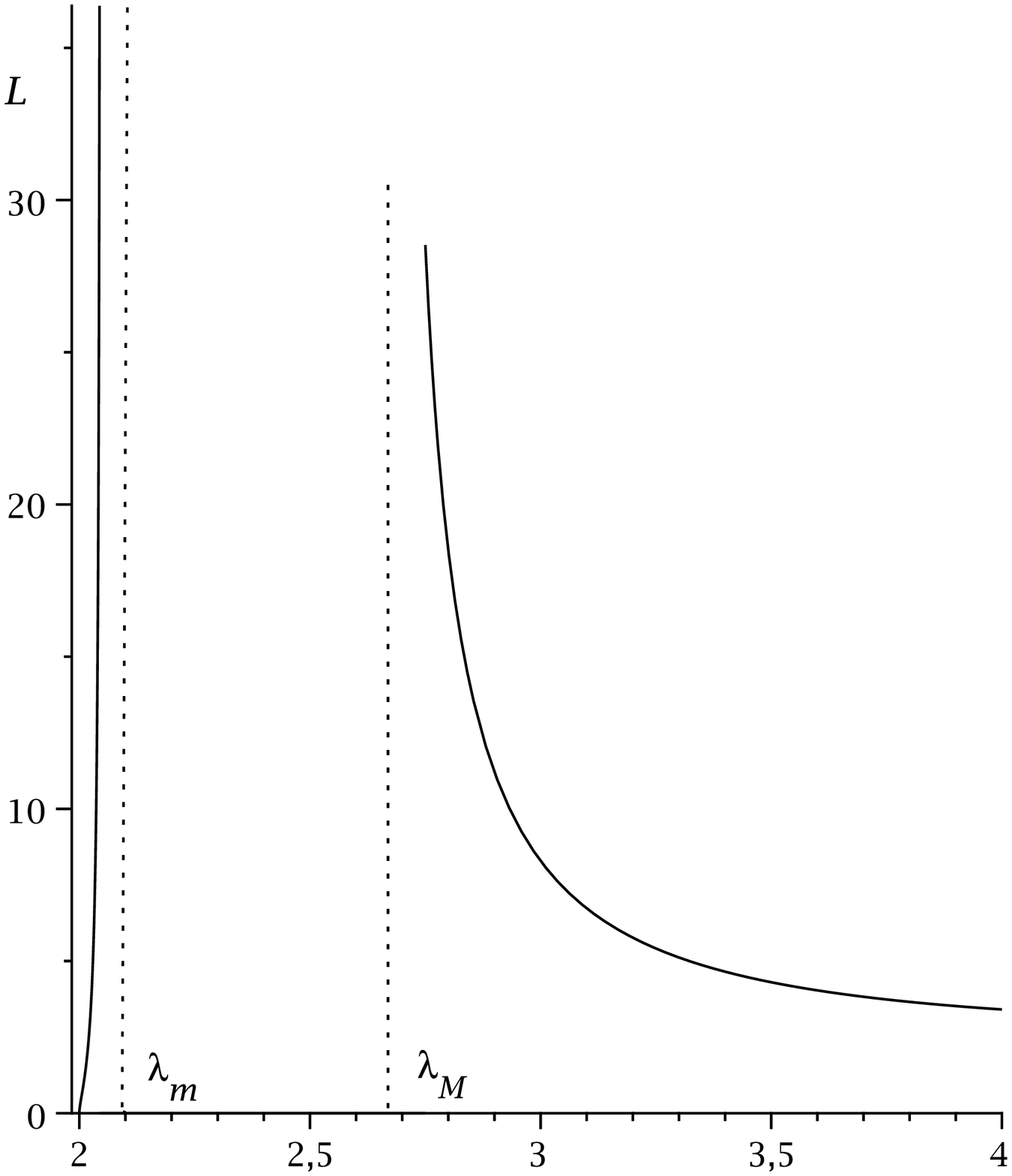} \nonumber
\end{array}
\nonumber
$$
\end{figure}

Such a behaviour is displayed in  figure 4 using the 
 equation in (\ref{lyh}), for different values of the quadrupolar parameter
$q$,  for the case of the M-Q$^{(1)}$ solution. For the discussion below  an
important role will be played by the function  $g(r)\equiv
1-2r\Psi^{\prime}$, which is related to $L$ by 
\begin{equation}
\frac{d L}{d r}= \frac{1}{4M} \frac{r^3 \Psi^{\prime \prime}+3
r^2\Psi^{\prime}+4r^4 (\Psi^{\prime})^3-6r^3 (\Psi^{\prime})^2}{e^{2\Psi}
g(r)^2}.
\end{equation}
The plot of $g$ as function of $\lambda$ is given in figure 5 for the
M-Q$^{(1)}$ and the LM solutions.

The following conclusions emerge from figures 4 and 5: 

\begin{itemize}
\item First, we see that stable orbits are located  to the 
left of the maximum, and   the right of the
minimum  (figure 4 (a)) where $L^{\prime}>0$. The slope of the curve where 
$L^{\prime}<0$ determines the range of the orbital radius for unstable
orbits. The mso is located at the maximum or minimum of the curve  ($L_+$ and
$L_-$ respectively). The values of the orbital radius ($mso_+$ and  $mso_-$ for
the maximum and minimum respectively) at these two extremals of $L$
correspond to the inflection points of the potential $-F$  for which this
function does not possess extremal points. For other values of $L \in
\left(L_+,L_-\right)$ (at each value of $q$) the potential $-F$ will
possess one maximum and two minima corresponding to the intersection points of
the curves in Figure 4(a) with the horizontal lines $L=cte$.

\item Second, figure 4(b) represents the curve for a value of the quadrupolar
parameter $q$ where the inner unstable orbit is limited by the asymptotic dashed
line, and the inner stable orbit is located at the minimum of the curve. This
plot recovers the behaviour of the spherical case but slightly modifying the
position of the circular orbits when a quadrupole  moment is present. The
relevant difference with respect to the spherical case is that
a maximum of $L$ ($L_+$)  arises  at the value of the radial coordinate
$mso_-$ whenever some multipole of the solution, higher than the monopole, is
not zero  (within a determined range of that multipole parameter), and
this fact leads to the existence of stable circular orbits at a 
radius smaller than those where the other known minima and the maximum appear.
In addition, these new stable circular orbits possess smaller
values of $L$ and energy $h$ than the former ones.

\item Finally, we notice the existence of a splitting (see the figure 4(c)) in
the admissible region of circular
orbits radius for some values of the multipole moments of our
 space-time. This fact was already discussed in
\cite{dis14},  where by means of numerical methods the authors obtain  results
that suggest the
existence of disconnected non-plunging regions at small radii. The existence of
such
regions could be tested, for instance, in the presence of accretion disks
forming a ring structure around the source. The analytical determination of that
region
consists of calculating the zeros of the function $g(r)$.

Thus, for some values of $q$ the maximum of $L$ disappears and a region
of forbidden circular orbits arises as is shown in figure 4 (c). That region
corresponds to the range of values of  the radial coordinate  leading to
$g(r)<0$ (let us remember that $L \geq 0$, $ h^2 \geq 0$).
 The zeros of the function $g(r)$ (see figure 5) provide the asymptotical
behaviour of $L$ at the values $\lambda_m$ and $\lambda_M$ (figure 4(c)).

\end{itemize}

To complement the discussion above it is instructive to take a look at tables I
and  II, which  display the  values of different parameters characterizing
ISCO's for the  LM solution with $g=1$ (M-Q$^{(1)}$). As mentioned before, if 
$q \in \left(0,\frac{4}{15}\right]$ the behaviour of circular orbits is similar
to the spherically symmetric case: the minimum of $L$ (graphic (b) in  figure
(4) defines the change from stable to unstable orbits, and thereby the minimal
value of the radius of the stable circular orbit ($mso_-$). 

The maximum of the orbital radius for the unstable orbit exhibits  (as in the 
Schwarzschild case) an asymptote in the value of  $\lambda$ which is smaller
than the corresponding to the spherically symmetric case ($\lambda=3$). 

There exists a critical value for $q$ ($q_c$) beyond which a gap in the range of
possible values of the radius of the circular orbit appears, for which there are
not  ISCO's. This is clearly indicated in the graphic (c) of figure 4, where the
gap is determined by the interval  $\left(\lambda_m, \lambda_M\right)$  for  $q
\in \left(\frac{4}{15}, q_c\right]$.

In the interval   $\left(0, \lambda_m\right)$ there are  ISCO's close to the
 infinite redshift surface. 

For the  M-Q$^{(1)}$ solution $q_c$ is given by:
\begin{equation}
q_c=0.373434, \,  r_s/M=   2.367 \ ,
\end{equation}
that corresponds  to the value of $q$ for which  $g(r)$ has a single zero (see
figure 5). 

If $q \in \left(q_c,\frac{8}{15}\right]$ there are  ISCO's in the interval 
$\left(0,L_+\right)$, with values of the orbital radius smaller than those
corresponding  to stable circular orbits for  $L_-$ where $L$ has a
minimum ($mso_-$).

Three comments are in order at this point:
\begin{itemize}
\item It should be stressed that the range of admissible values  of  the angular
momentum  ($0, L_+$) is quite large.
Therefore, ISCO's correspond to test particles with a wide range of angular
velocities.
\item  The energies corresponding to ISCO's are smaller than those corresponding
to larger values of the orbital radii.

 \item It should be observed that for the M-Q$^{(1)}$ solution there are ISCO's
(inner to $3M$) only for positive values of $q$ (i.e. prolate sources). This
important difference between both cases (prolate and oblate) has been brought
out before for the $\gamma$ \cite{HS} and the M-Q$^{(1)}$ \cite{Herrera1, lhn2},
spacetimes. We ignore what could be (if any) the fundamental physical reason for
such a difference.
\end{itemize}

\begin{figure}[h]
\caption{\label{ges}\it  Plot of $g$ as function of $\lambda$ for  the 
M-Q$^{(1)}$ (a) and LM  (b) solutions   for different values of the multipole
parameters. The solid line in both plots corresponds to the Schwarzschild case
$q=0$, whereas the other values of $q$  are: (a) upper dottted line $q=0.5$, dashed
line $q=0.374\sim q_c$ and from that line  and downwards $q=0.3, 4/15, 0.2$, The
special value $q=4/15$ corresponds to the lower bound for the existence of ISCO
near the horizon. (b) from the solid line  $q=m_4=0$  and upwards,
$q=m_4=-0.063, -0.08, -0.1, -0.124$.}
$$
\begin{array}{cc}
\includegraphics[scale=0.20]{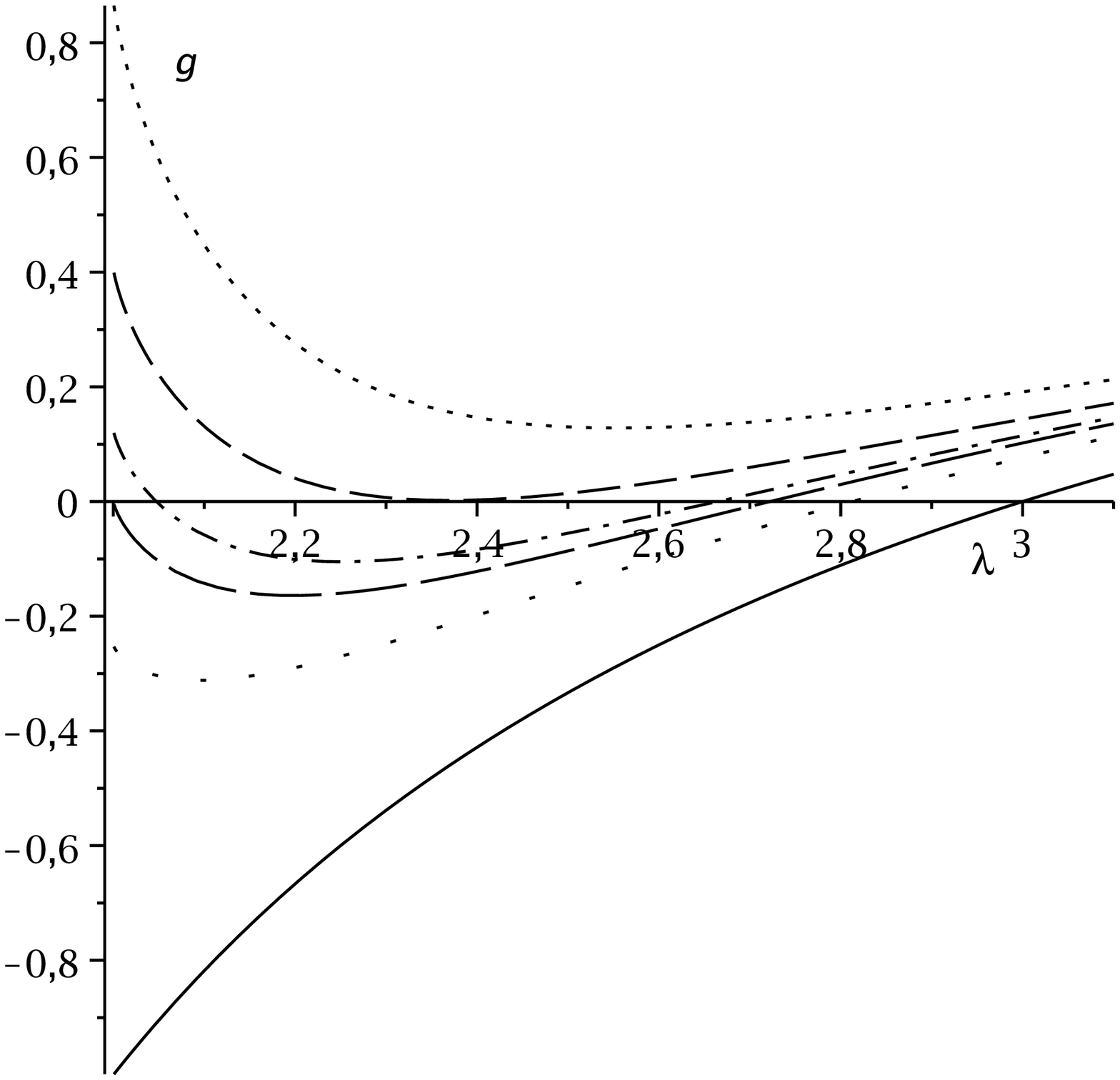} &
\includegraphics[scale=0.20]{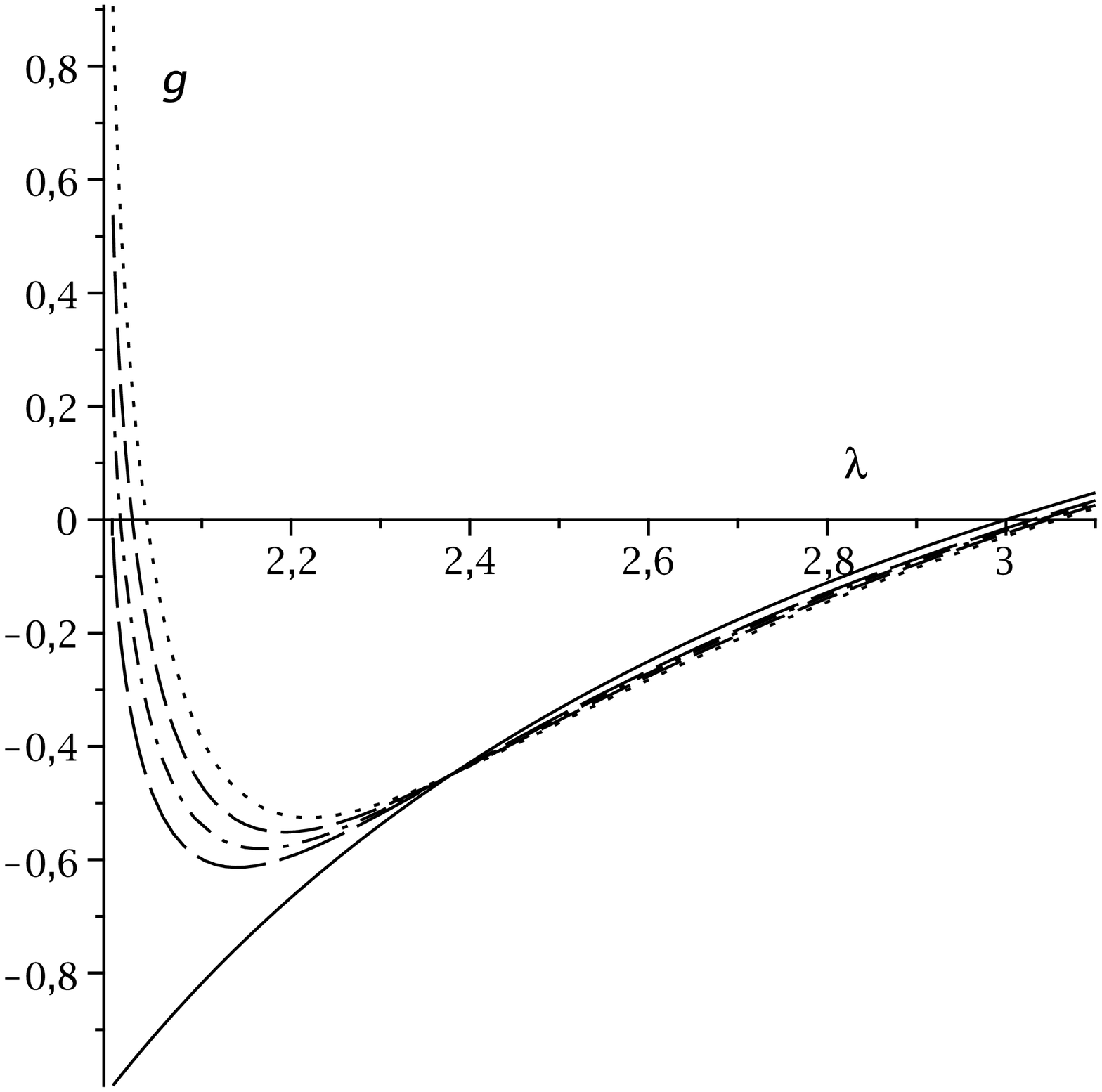}\\ \nonumber
(a) & (b) \nonumber
\end{array}
$$
\end{figure}

  Finally,  table III displays some values of relevant parameters ($mso_-$,
$L_-$ and $\lambda$), as well as the interval of non existence of stable
circular orbits ($\lambda_m$, $\lambda_M$), for the LM solution with quadrupole
and  $2^4$-pole. 

It should be observed that now, unlike the case of the  M-Q$^{(1)}$ solution,
the function  $L (\lambda)$ has no maximal value, implying there exists no 
$mso_+$. Thus, the existence of   ISCO's is restricted to the interval  $r_s/M
\in (0, \lambda_m)$ whenever the quadrupole and the  $2^4$-pole are localized 
within the range mentioned before  (\ref{iscom4q}). 

Also, the value of  $\lambda_m$ is significantly reduced with respect to the
M-Q$^{(1)}$  case, and therefore  ISCO's s are now very close to the infinite redshift surface 
($r_s/M=2$). At the same time, the range  ($\lambda_m$, $\lambda_M$) increases
with respect to the previous case. For values of  $r_s/M$  starting from 
$mso_-$ (minimum of  $L$) we obtain the values of the farthest possible
stable circular orbits.

\section{Conclusions}
We have presented a systematic study on the structure of circular geodesics in
the LM spacetime. The case has been made for the use of such spacetime when
describing slight deviations from spherical symmetry.

The analysis presented clearly exhibits the difference between the motion in the
Schwarzschild  and  in the LM, spacetimes. In the former case we have a black
hole whereas in the latter a naked singularity appears. Our results, as well as
those in the  references already mentioned, point to a potentially observable
evidence allowing  to distinguish between the two above mentioned situations. 
We may summarize such results as follows:
\begin{itemize}
\item The presence of multipole moments (higher than the monopole) leads to the
presence of ISCO's, closer to the infinite redshift surface  than the ones existing  in the exactly
spherically symmetric case.
\item Such multipole moments also produce an interval in the values of radial
coordinate within which no stable circular orbits exist. 
\item Specific numerical values have been presented to illustrate the two
abovementioned effects.
\item Particularly relevant  might be the application of  the presented results
to studying the dynamics of accretion discs around compact objects, which as it is
well known, are assumed to be an essential ingredient  of active sources such as
$X$-ray binaries or galactic  nuclei (see \cite{jap} and references therein).
However such a study is out of  the scope of this paper.
\end{itemize}

\section{Acknowledgments}
This  work  was partially supported by the Spanish  Ministerio de Ciencia e
Innovaci\'on under Research Project No. FIS 2012-30926, and the Consejer\'\i a
de
Educaci\'on of the Junta de Castilla y Le\'on under the Research Project Grupo
de Excelencia GR234.

\begin{table}[h]
\caption{\label{tabI} Numerical values of  characteristic parameters
corresponding to  ISCO's in the  M-Q$^{(1)}$ solution.}
\begin{ruledtabular}
\begin{tabular}{cccc}
$q$ & $L$ & $h^2$& ISCO($r_s/M$) \\
0 & 3 &0.889 & 6\\
0.28& 0.1 & 0.016 & 2.000\\
0.28& 1.1 & 0.147 & 2.001\\
0.28 & 2.1 & 0.269 & 2.004\\
0.28 & 3.1 & 0.388 & 2.006\\
0.28 & 4.1 & 0.508 & 2.007\\
0.28 & 5.1 & 0.626 & 2.008\\
0.34& 0.1 & 0.053 & 2.003\\
0.34& 1.1 & 0.259 & 2.041\\
0.34 & 2.1 & 0.433 & 2.062\\
0.34 & 3.1 & 0.602 & 2.076\\
0.34 & 4.1 & 0.769 & 2.086\\
0.34 & 5.1 & 0.935 & 2.093\\
0.40& 0.1 & 0.093 & 2.015\\
0.40& 1.1 & 0.335& 2.104\\
0.40 & 2.1 & 0.535 & 2.148\\
0.40 & 3.1 & 0.728 & 2.179\\
0.40 & 4.1 & 0.918 & 2.202\\
0.40 & 5.1 & 1.106 & 2.222\\
0.46& 0.1 & 0.130 & 2.038\\
0.46& 1.1 & 0.392 & 2.174\\
0.46 & 2.1 & 0.606 & 2.245\\
0.46 & 3.1 & 0.811 & 2.301\\
0.46 & 4.1 & 1.010 & 2.352\\
0.46 & 5.1 & 1.202 & 2.407\\
0.52& 0.1 & 0.163 & 2.068\\
0.52& 1.1 & 0.436 & 2.248\\
0.52 & 2.1 & 0.657 & 2.350\\
0.52 & 3.1 & 0.865 & 2.446\\
0.52 & 4.1 & 1.064 & 2.577\\
\end{tabular}
\end{ruledtabular}
\end{table}
\newpage

\begin{table}[h]
\caption{\label{tabII} Extremal values of the marginally stable orbits  $mso_+$
and $mso_-$ with the corresponding value of the angular momentum parameter
$L$ for which ISCO's exist, and the range of non existence of stable orbits,
for different values of the quadrupole moment, for the  M-Q$^{(1)}$ solution.}
\begin{ruledtabular}
\begin{tabular}{ccccccc}
$q$ & $mso_+$ & $L_+$ &$ mso_-$ & $L_-$ & $(\lambda_m,\lambda_M)$ &
$\lambda$  \\
0 & - & -& 6&3&-&3\\
0.04 & - & -&5.957 &2.987&-&2.968\\
0.08& - & -& 5.913&2.973&-&2.934\\
0.12 & - & -& 5.868&2.960&-&2.898\\
0.16 & - & -& 5.821&2.946&-&2.857\\
0.20 & - & -& 5.772&2.931&-&2.812\\
0.24 & - & -& 5.722&2.917&-&2.761\\
0.26 & - & -& 5.696&2.909&-&2.732\\
0.28 & - & - &5.670 &2.902 & 2.014, 2.701& -\\
0.30& - &  -&5.643 & 2.894 & 2.047, 2.665& -\\
0.32& - & -& 5.615 & 2.886& 2.092, 2.623& -\\
0.34& - & - &5.587 & 2.878& 2.149, 2.571& -\\
0.36& - & - &5.559 & 2.870&2.230, 2.498&-\\
0.37& - & - &5.544  &2.866  &2.298, 2.434&-\\
0.374&2.374&7.04e$^9$&5.538 &2.865   & - &-\\
0.40&2.435&18.277&5.500&2.854&-&-\\
0.44&2.540&8.152&5.436&2.837&-&-\\
0.48&2.645&5.634&5.370&2.820&-&-\\
0.52&2.753&4.498&5.298&2.801&-&-\\
8/15&2.789&4.248&5.273&2.795&-&-\\
\end{tabular}
\end{ruledtabular}
\end{table}

\begin{table}[h]
\caption{\label{tabIII} The case of LM solution with $g=2$, i.e, quadrupole and
$2^4$-pole moments. Numerical values of the multipole parameters $q$ and $m_4$
for which  ISCO's exist. The value of the marginally stable orbit $mso$ is given
 with the corresponding value of the angular momentum parameter $L$. }
\begin{ruledtabular}
\begin{tabular}{ccccc}
$q$  &$m_4$ &$ mso_-$ & $L_-$ & $(\lambda_m,\lambda_M)$   \\
-0.063 & -0.063&6.000 &3.0195 & 2.000, 3.030\\
-0.08 &-0.08 &6.000 & 3.0248 &2.0091,  3.038\\
-0.10 & -0.10&6.000 & 3.0311 & 2.0224, 3.047\\
-0.124 & -0.124& 6.000&3.0386 & 2.0385, 3.058\\
0.0427 & -0.0427&5.9501 &2.9852 & 2.000, 2.953\\
0.05 &-0.05&5.9414& 2.9826&2.0081,  2.943\\
0.06 & -0.06&5.9294 & 2.9791 & 2.024, 2.931\\
0.07 & -0.07& 5.9173&2.9756& 2.044, 2.917\\
0.085 & -0.085& 5.8989&2.9702 & 2.075, 2.899\\
\end{tabular}
\end{ruledtabular}
\end{table}

\end{document}